**Planar Microcavities can Suppress Exciplex Formation and Increase the Emission Efficiency of Organic Semiconductors**

Tomohiro Ishii[1, 2 †] and Stéphane Kéna-Cohen[1*]

[1] Department of Engineering Physics, École Polytechnique de Montréal, Montréal H3C 3A7, QC, Canada.

[2] Research Institute for Electronic Science, Hokkaido University, Sapporo 001-0020, Japan

E-mails: [†]t.ishii@es.hokudai.ac.jp, [*]s.kena-cohen@polymtl.ca

**Abstract**

**Optical microcavities are widely used to control the emission of organic semiconductors, but their ability to reshape the molecular pathways that precede emission remains largely unexplored. Here we show that embedding a ZnPc:TPBi blend in a planar Fabry-Pérot microcavity suppresses the formation of non-radiative exciplexes and removes bimolecular annihilation at high excitation densities, increasing the photoluminescence quantum yield by more than forty-fold under continuous-wave excitation. This enhancement is far larger than expected from the weak Purcell effect. Instead, transient spectroscopy, power-dependent photoluminescence and kinetic modelling point to a cavity-induced rebalancing of excited-state populations: long-range Förster energy transfer from ZnPc monomers to emissive aggregates is enhanced, allowing it to outcompete charge transfer to dark exciplexes. Electromagnetic calculations predict FRET enhancements of up to ~400-fold at relevant distances, consistent with the observed suppression of exciplex-mediated losses. Our results show that optical cavities can control not only how molecules emit, but also which excited states they form, opening a route to improved efficiency and reduced roll-off in organic optoelectronic and photonic devices.**

## ・ Introduction

Organic semiconductors have attracted considerable attention for applications in optoelectronic devices such as organic light-emitting diodes (OLEDs), organic solar cells, and organic lasers, owing to their flexibility, low weight, and compatibility with low-cost fabrication processes[1-7]. The emission efficiency of organic molecules in these devices is largely determined by the relaxation pathways of the various excited states, in particular by processes such as energy transfer and charge transfer. Among them, exciplexes, which are excited-state complexes formed between pairs of donor and acceptor-like molecules, are prone to non-radiative decay via internal conversion and intersystem crossing from singlet to triplets [8-11]. Exciplex formation can significantly reduce the photoluminescence quantum yield (PLQY) of emissive layers and often impose fundamental limitations on organic semiconductor device performance[12-16]. In this work, we demonstrate the suppression of exciplex formation and a resulting enhancement of the emission efficiency of a molecular blend by embedding it within a Fabry–Pérot microcavity.

In recent years, theoretical and experimental efforts have focused on exploring how light–matter interactions in optical microcavities can modify molecular kinetics beyond the Purcell effect[17-31]. It is well known that in microcavities, modifications of the photon local density of states lead to changes in the radiative decay rate as well as the outcoupling efficiency[17,18,25,30,32-35]. Less explored is the fact that in a cavity environment, the efficiency of Förster resonance energy transfer (FRET) is also modified. Cavities can modify FRET rates, which are proportional to the donor electric field at the acceptor position, by more than an order of magnitude for molecules placed tens of nanometers apart[36-39]. However, these modifications are often neglected because the electric near field, which decays as $R^{-6}$, generally dominates FRET in undoped films. This near-field contribution is nearly independent of the optical environment. Cavity-induced effects only play an important role if the average donor-acceptor pair distance reaches the intermediate-field length scale.

To explore this mechanism, we focus on zinc phthalocyanine (ZnPc; molecular structure shown in Fig. S1.1a). ZnPc is a widely used organic optoelectronic material owing to its strong optical absorption, chemical stability, and structural tunability via molecular design[40-43]. In the solid state, ZnPc tends to undergo $\pi$–$\pi$ stacking, leading to the formation of H-type aggregates[44], which are known to possess intramolecular charge-transfer (CT) character in the excited state[45-48]. As an electron donor, ZnPc readily forms charge transfer states at interfaces with electron-accepting molecules[49-51].

When doped at 50 wt% in 2,2′,2"-(1,3,5-Benzinetriyl)-tris(1-phenyl 1-H-benzimidazole) (TPBi), ultrafast charge transfer from ZnPc monomers to TPBi (<200 fs) leads to exciplex formation ($ZnPc^{+}$-$TPBi^{-}$), while FRET from ZnPc monomers to ZnPc aggregates ( $\tau_{FRET}$ = 1.09 ± 0.09 ps) competes with exciplex formation (Fig. 1a). In ZnPc:TPBi, exciplexes undergo fast non-radiative decay, resulting in a PLQY for 50 wt% TPBi:ZnPc films of only (2.8 ± 0.5) $\times 10^{-2}$ %.

We will show that modifying the balance between these processes via the optical environment can have a profound impact on the overall molecular kinetics and consequently the film PLQY. In microcavities, FRET from ZnPc monomers to ZnPc aggregates is enhanced, but only on a >10 nm length scale (Fig. 1b). When the excitation density is high enough such that nearby aggregates are occupied, the contribution from long-range FRET begins to dominate. This can be seen as a systematic increase in PLQY with increasing exciton density, reaching up to ~5-fold compared to the control sample in the linear regime (Figs. 1c and 1d). Furthermore, the suppression of exciplex formation nearly eliminates singlet–triplet annihilation, which is the dominant bimolecular annihilation mechanism in ZnPc:TPBi blends. Together, these effects lead to a >40-fold increase in the microcavity PLQY at the highest excitation densities (Fig.

1d). These results highlight a novel strategy for improving the emission efficiency and PLQY roll-off of organic semiconductors by controlling the balance of excited states through modifications of the FRET rate via the optical environment.

**Optical and structural analysis of the ZnPc thin films**

Understanding the photophysics of the TPBi:ZnPc blends is paramount if one hopes to decipher the mechanisms at play in the microcavities. The photoluminescence and absorption of 2 wt% and 50 wt% TPBi:ZnPc films are shown in Fig. 2a (see Fig. S1.1 for neat ZnPc). In the ZnPc 2 wt% film (Fig. 2a), narrow Q-band transitions are observed in absorption at 685 nm and 620 nm, with a full width at half maximum (FWHM) of 26 nm, corresponding to the ZnPc monomer[52]. The corresponding PL spectrum has a mirror-symmetric shape with a small Stokes shift (~ 6.3 nm). In contrast, neat ZnPc shows broad absorption bands centered at 625 nm and 713 nm, with a significantly larger FWHM of ~72 nm (Fig. S1.1b). The PL spectrum of neat ZnPc is centered at 986 nm with a FWHM of 116 nm, consistent with emission originating from excimers and H-aggregates[44,53,54]. Similarly, the TPBi:ZnPc 50 wt% blend film shows absorption bands at 632 nm and 692 nm, both with FWHMs of ~38 nm (Fig. 2a). Notably, the lower-energy peak at 632 nm is more intense than the higher-energy peak at 692 nm, suggesting the coexistence of monomer and aggregated ZnPc, a conclusion further supported by the transient absorption (TA) data discussed below (Fig. 2b). Furthermore, the PL spectrum of the 50 wt% film is broad and centered at 938 nm with an FWHM of 144 nm, indicating that, as in the neat ZnPc film, the emission originates from excimers and ZnPc aggregates[44,53,54]. X-ray diffraction for the neat ZnPc film confirmed that it was polycrystalline and composed of monoclinic α-phase domains[55] (see Fig. S1.3). In contrast, the TPBi: ZnPc 50wt% film was amorphous, showing no diffraction peaks originating from the $\alpha$-phase.

**Förster energy transfer from monomers to aggregates in ZnPc thin films**

To better understand the excited state kinetics of the various films, pump–probe transient absorption results are shown in Fig. 2b. In the 200 nm-thick 50 wt% ZnPc film, a ground state bleach (GSB) signal at 680 nm is observed within 0.1 ps after photoexcitation (Fig. 2c). This GSB spectrum is consistent with that of the TPBi: ZnPc 2 wt% film (Fig. S1.9), and we ascribe it to the excitation of ZnPc monomers within the film. Within 1 ps, however, the monomer-derived GSB is suppressed, and a GSB signal emerges at ~623 nm as shown in Fig. 2c. By comparing with the absorption spectrum, we can ascribe this feature to ZnPc aggregates, and the TA kinetics indicate highly efficient FRET from ZnPc monomers to aggregates. The characteristic time constant is found to be $\tau_{FRET} = 1.09 \pm 0.09$ ps from global fitting. The wavelength-resolved kinetics further support this: as the monomer-derived GSB (680 nm) decays, a concomitant growth of ZnPc aggregate excited-state absorption (ESA) around 550 nm[56,57] and aggregate GSB are observed (Fig. 2d). This ESA signature is also consistent with density functional theory calculations reported for ZnPc monomers, dimers, and trimers[56,57].

Global analysis of the TA data revealed three spectrally and temporally distinct excited-state species: monomer singlets, aggregate singlets, and aggregate triplets (Fig. S1.4). This analysis further yields the intersystem crossing time constant ($\tau_{ISC}$) of $0.15 \pm 0.05$ ns for the conversion from singlet to triplets in ZnPc aggregates (Fig. 2e). This value is similar to that found in ZnPc neat films, in agreement with the aggregate triplet assignment (Fig. S1.13).

**$ZnPc^{+}$-$TPBi^{-}$ exciplexes in the TPBi:ZnPc 50 wt% film**

To understand the role of host-matrix interactions in TPBi:ZnPc (50 wt%) films, we investigated the properties of a PMMA:ZnPc film in which host–guest interactions are absent. The absorption spectrum of the PMMA:ZnPc film shows a broad absorption band at ~615 nm and a narrow transition at 670 nm, similar to those observed in the neat ZnPc and TPBi:ZnPc (2 wt%) films (Fig. S1.1 and Fig. 2a), indicating that ZnPc monomers and aggregates coexist in the PMMA blend. In addition, the PL spectrum shows a broad emission band centered at ~800 nm with a FWHM of 48 nm, attributed to ZnPc aggregates, as well as a small emission peak at 674 nm arising from ZnPc monomers (Fig. S1.2).

The TA spectrum at 0.1 ps for the PMMA:ZnPc film showed an ESA signature near 550 nm (Fig. S1.7), similar to the other films. In contrast, ESA above 765 nm was completely absent in PMMA:ZnPc (Fig. S1.7). This long-wavelength ESA, which is present in the TPBi:ZnPc 2wt% and 50 wt% films (Fig. S1.5), points to the formation of an excited species resulting from TPBi–ZnPc interactions. Furthermore, the PLQY of the TPBi:ZnPc 50 wt% film was estimated to be $(2.8 \pm 0.5) \times 10^{-2}$ %, which is markedly lower than that of PMMA:ZnPc, for which the PLQY = 0.13 ± 0.01 %. The reduction of PLQY and the ESA signal above 765 nm are both ascribed to the formation of a non-radiative $ZnPc^{+}$–$TPBi^{-}$ exciplex.

In Fig. 2d, we can observe ultrafast charge transfer from ZnPc to TPBi within the instrument response function (IRF, <0.2 ps), evidenced by the simultaneous rise of the ZnPc GSB and exciplex ESA signals. At longer timescales (~1.5 ns), ESA associated with triplet exciplexes is also observed (Fig. S1.6). This notably overlaps with the emission spectrum of the TPBi: ZnPc 50 wt% film (Fig. 2a), which we will see later is a source of singlet-triplet quenching. An energy-level diagram summarizing the various decay pathways in TPBi:ZnPc (50wt%) is shown in Fig. 2f.

**Enhanced PLQY in microcavities and suppressed singlet-triplet annihilation**

To investigate how microcavities affect the relaxation pathways in TPBi: ZnPc (50 wt%) films, we fabricated a series of weakly coupled microcavities with different cavity thicknesses (Fig. 3a), corresponding to normal-incidence cavity mode wavelengths of $\lambda_c$ = 780, 860, 880, and 920 nm. Angle-resolved PL spectrum for the microcavity with $\lambda_c$ = 880 nm are shown in Fig. 3b. We measured the photoluminescence quantum yield (PLQY) and excitation power dependence of the PL for these microcavities, as well as for identical control samples deposited on fused silica and silver substrates (Figs. 1d and 3c). The TPBi: ZnPc (50 wt%) layer thickness was fixed at ~5 nm to ensure that the entire film was located at the electric-field antinode while maintaining the system in the weak-coupling regime. Further details of the sample fabrication and measurements are provided in the Methods section and Supplementary Information Sections 2.1–2.2.

Figure 1d shows the PLQYs of the fused silica control samples as a function of the absorbed power density under CW laser excitation (blue open squares). At high excitation densities, the PLQY decreases by nearly an order of magnitude to $(3.7 \pm 0.1) \times 10^{-3}$ % due to bimolecular annihilation. In contrast, Fig. 1d (red open squares) shows the power dependence of the PLQY for the microcavity with $\lambda_c = 880$ nm, extracted from quantitative *k*-space PL measurements (Figs. S2.15). The PLQY of the microcavity increases systematically from 0.029% to 0.12% when the absorbed pump intensity is raised from 3.75 W/cm$^2$ to 150 W/cm$^2$. Figure 3c and Figure S2.11 compares the relative PL intensity of the microcavities ($\lambda_c = 880$ nm) with that of the control sample. Although the control sample shows sublinear behavior above 20 W/cm$^2$, the PL of the microcavity remains linear up to 2 kW/cm$^2$. Compared with the control sample, these observations correspond to a ~5-fold enhancement of the PLQY at low power and a >40-fold enhancement at high power, reflecting both enhanced radiative aggregate formation and the suppression of annihilation-induced roll-off. Similar behavior is observed for the other microcavities ($\lambda_c$ = 780, 860, and 920 nm), where the PLQYs at absorbed intensities of ~150 W/cm$^2$ are enhanced by factors of 2.2, 4.3, and 3.3, respectively compared to the control at low power (see Supplementary information 2.7.2).

Several observations rule out a conventional Purcell or outcoupling explanation. The reported PLQYs are already corrected for the modified outcoupling efficiency of the microcavities (see Table S2.8), so the enhancement reflects changes in internal excited-state kinetics rather than improved photon extraction. In addition, the broad ZnPc:TPBi emission strongly reduces the relevant Purcell enhancement: when averaged over the measured emission spectrum, the calculated value is close to unity ($F_p = 1.08$) for the $\lambda_c$ = 880 nm cavity (see Table S2.15). Finally, under low-repetition-rate pulsed excitation the microcavity PL becomes comparable to that of the control sample (Fig. 1d). These observations point to a cavity-induced change in the excited-state population dynamics preceding emission, with long-lived states playing an important role.

**Time-resolved PL characteristics in control samples and microcavities**

To understand how microcavities can modify the kinetics of the TPBi: ZnPc 50 wt% films, we performed time-resolved photoluminescence measurements of the $\lambda_c$ = 880 nm microcavity and the corresponding control sample on fused silica. As shown in Fig. 3d, the prompt PL lifetime is systematically shortened with increasing excitation power. This quenching behavior depends on the average pump power, as opposed to the pulse energy, and vanishes in low repetition rate experiments where triplets are absent. These observations indicate that quenching originates from singlet-triplet annihilation (STA). Because of the strong spectral overlap between the aggregate PL and the ESA of the long-lived exciplex triplets (see Fig. S1.6 and Fig. 2a), we can assign the microscopic origin to aggregate singlet excitons and exciplex triplets.

Figure 3e shows the microcavity transient response measured under the same absorbed power conditions as the control sample. In contrast to the control sample, the microcavity PL lifetimes hardly change across the range of powers (Fig. S2.5). In fact, the delayed fluorescence systematically increases with excitation density, following a trend completely opposite to that of the control film.

To quantitatively understand the photophysics of the control fims, we use a rate equation model that includes all of the relevant excited states and the various rates obtained from TA and time-resolved spectroscopy (see Supplementary Information S2.4). The model succesfully reproduces all of the features observed in experiment, including the measured PLQY and the PL (Fig. 1d and Fig. 3c) and transient PL power dependences (Fig. S2.9).

**Suppressed exciplex formation and accelerated long-range FRET in microcavities**

The increased PLQY (~ 0.07 – 0.12 %) and the absence of STA in the microcavities are in good agreement with the photophysics observed in PMMA:ZnPc film (PLQY = 0.13 ± 0.01%) in which exciplex formation is absent. This suggests that the modified optical environment is somehow suppressing exciplex formation.

Having ruled out a dominant contribution from outcoupling or Purcell enhancement, we postulate that the suppression of exciplex-mediated losses arises from a cavity-induced modification of the FRET rate between initially excited ZnPc monomers and emissive ZnPc aggregates (Fig. 1b). To quantitatively evaluate this possibility, we calculate the FRET rate using a dyadic Green's function approach.[58] The calculation is performed separately for donor transition dipole moments oriented perpendicular ($z$-direction) and parallel ($x$-direction) to the cavity plane (see Fig. 4c), while acceptor dipoles are assumed to be randomly oriented. Figures 4a and 4b show the calculated FRET enhancement, as compared to a homogeneous environment, for the $\lambda = 880$ nm cavity as a function of in-plane donor–acceptor distance ($R_x$), and wavelength. For $z$-oriented dipoles, we find that FRET is enhanced by up to ~ 400 in the 15–100 nm range, whereas for $x$-oriented dipoles, an enhancement of up to ~50-fold is found. These results show that while FRET over distances of a few nanometres, which typically dominates in neat films due to its $1/R^6$ dependence, remains unchanged, the cavity environment can strongly modify FRET rates in situations where average donor-acceptor separations exceed ~10 nm.

The maximum FRET enhancement is calculated to occur near $\lambda = 690$ nm, which coincides with the ZnPc monomer emission and ZnPc aggregate absorption (Fig. 2a). This supports the hypothesis that cavities can increase the FRET rate from ZnPc monomers to aggregates, $k_{FRET}$, which was observed in TA to compete with the rate of exciplex formation, $k_{CT}$. The fact that the microcavity enhancement appears only under CW or high-repetition-rate excitation provides an important clue to the underlying mechanism. At low excitation density, the earliest dynamics are dominated by local processes: ultrafast charge transfer to TPBi and short-range FRET to nearby ZnPc aggregates. By contrast, long-range FRET is intrinsically slower because the donor–acceptor separation is larger, even when its rate is enhanced by the cavity. It therefore becomes relevant only once the local relaxation channels are partially saturated by accumulated long-lived states. Under CW excitation, aggregate and exciplex triplets can occupy nearby aggregate- and CT-forming sites, reducing the effective rates of both short-range FRET and charge transfer. This blockade effect alone is not sufficient to improve the PLQY, since it is also present in the control film. The key difference in the microcavity is that, under these same blocked conditions, the optical environment strongly enhances the long-range FRET channel (Fig. 4d). This provides an additional pathway for ZnPc monomer excitations to reach emissive aggregates rather than forming non-radiative exciplexes, explaining both the increase in PLQY and the suppression of STA under CW excitation. Our simulations show that at the absorbed powers where these effects play role, CT and aggregate triplet densities are of order ~$10^{19}$–$10^{20}$ $cm^{-3}$ (Fig. S2.7), which is comparable to expected number densities for exciplex and aggregate-forming sites.

This picture can be incorporated into the rate-equation model by treating the cavity as an enhancement of the effective FRET rate from ZnPc monomers to aggregates. In reality, blockade by accumulated long-lived triplets is expected to introduce density-dependent changes in both the effective FRET rate and the charge-transfer rates in the model. To avoid overfitting, we do not explicitly include these separate blockade terms. Figure S2.8 and Table S2.4 shows the FRET enhancement required in the kinetic model to reproduce the microcavity

PLQY. This value increases with pump power, consistent with a growing contribution from cavity-enhanced long-range FRET as nearby relaxation pathways become saturated. Meanwhile, an enhancement factor of 35 is sufficient to suppress singlet–triplet quenching and reproduce the power dependence observed experimentally, as shown in Fig. 1d and Fig. 3c. (see Suppementary Information 2.4 for additional results).

**Conclusions**

In this work, we show that a structured optical environment can be used to control the fate of molecular excited states before emission occurs. In ZnPc:TPBi blends, placing the active layer inside a planar Fabry–Pérot microcavity strongly reduces exciplex-mediated losses and suppresses bimolecular annihilation that otherwise limits the PLQY at high excitation densities. The resulting enhancement cannot be accounted for by changes in radiative decay alone, pointing instead to a redistribution of population among competing excited-state pathways. Our measurements and modelling support a picture in which, under conditions where accumulated long-lived states partially block local relaxation channels, cavity-enhanced long-range FRET increases the probability that ZnPc monomer excitations reach emissive aggregate states rather than forming non-radiative exciplexes. The calculated enhancement of this long-range transfer channel over tens of nanometres suggests that even weakly coupled planar cavities can influence molecular kinetics when the relevant processes occur on mesoscopic length scales. More broadly, these results expand the role of optical cavities in organic semiconductors from controlling emission to controlling excited-state branching, providing a route to reducing loss and roll-off in cavity-integrated optoelectronic and photonic devices.

・ **Methods**

**Sample fabrication**

The TPBi: ZnPc (2 wt% and 50 wt%) thin films, the ZnPc neat film, microcavities, and the control samples were fabricated on Si substrates by thermal evaporation at a base pressure below $10^{-7}$ mbar (EvoVac, Angstrom Engineering). NPB (CAS: 123847-85-8, >99.5%) and ZnPc (CAS: 14320-04-8, >99%) were purchased from Lumtec, and TPBi (CAS: 192198-85-9, 99.9%) from Xi'an Yuri Solar Co., Ltd.

The PMMA: ZnPc film was prepared by dissolving PMMA (34.1 mg) and ZnPc (0.344 mg) in dimethyl sulfoxide (3.1 mL), followed by ultrasonication to achieve a homogeneous dispersion. The resulting solution was drop-cast onto a fused silica substrate heated to 165 °C and subsequently dried for 30 minutes to form the film. The film thickness (~ 1400 nm) was determined using a Dektak profilometer (Bruker).

**Sample information for the ZnPc thin films**

We prepared the 200 nm-thick TPBi: ZnPc 2wt%, 200 nm-thick TPBi: ZnPc 50wt% thin films, the 50 nm-thick ZnPc neat film, and the 1400 nm-thick PMMA: ZnPc film on a fused silica substrate. To protect the samples from air exposure, they were encapsulated with CYTOP (200 nm) (CAS: 37626-13-4, AGC Chemicals).

**Sample characterization for ZnPc thin films**

X-ray diffraction measurements were carried out using a symmetric $2\theta$-$\theta$ measurement (Burker). Absorption measurements were carried out using a commercial spectrophotometer (PerkinElmer Lambda 25). PL, and PLQY measurements were performed with a fluorometer (HORIBA Scientific, Fluorolog-QM). Refractive indices and film thicknesses were determined using a variable-angle spectroscopic ellipsometer (J.A. Woollam Co., RC2 D + NIR).

Transient absorption (TA) measurements were performed using a pump–probe setup (Light Conversion, HARPIA-TA). An ultrafast optical parametric oscillator (Light Conversion,

Orpheus-F) provided pump pulses at $\lambda = 680$ nm, while a white-light continuum generated in a sapphire crystal was used as the probe pulse. The laser repetition rate was 100 kHz and the pulse fluence was 290 μJ/cm$^2$. Global analysis of the TA data was conducted using CarpetView (Light Conversion). All measurements were performed in the linear regime.

**Sample information for microcavities and control samples**

We prepared four types of microcavities with resonance peak positions at normal incidence of $\lambda_c = 780$ nm, 860 nm, 880 nm, and 920 nm. The structures consisted of Ag/TPBi/NPB/TPBi: ZnPc (50 wt%)/NPB/Ag (see Table S2.1). The thicknesses of each layer were obtained by fitting the normal-incidence reflectance spectra measured with a spectrophotometer (Cary 7000, Agilent) in combination with ellipsometry data (Fig. S2.3). The fraction of incident power absorbed within the ZnPc layer was calculated using the Transfer Matrix Method (see Supplementary Information 2.2). Simultaneously-grown control samples, such as TPBi (10.4 nm)/NPB (94 nm)/TPBi: ZnPc (50 wt%, 4 nm)/NPB (94 nm) were fabricated on both silver and fused silica substrates. Absorption was measured using a spectrophotometer (Cary 7000, Agilent).

**Sample characterization for Microcavities and control samples**

Angle-resolved reflectance measurements were performed using a variable-angle spectroscopic ellipsometer (J.A. Woollam Co., RC2 D + NIR) in Fig. S2.2.

Angle-resolved PL was measured by Fourier imaging with excitation from a continuous-wave HeNe laser at 633 nm (CW RADIATION, Inc.). Imaging was conducted on an Olympus IX-81 inverted microscope using a 100× objective lens (NA = 0.95, Olympus), with a laser spot size of $4.15 \times 10^{-9}$ cm$^2$. The emission was collected with an imaging spectrometer coupled to a cooled CCD camera (Princeton Instruments IsoPlane 160 spectrometer and Retiga R1™ CCD camera). Long-pass filters at 632.8 nm (Ultrasteep HeNe PN-ZX000548) and 700 nm (FELH0700, Thorlabs) were used to block the excitation laser before detection. The microcavities were excited with 200 μW incident power (corresponding to 0.623 μW absorbed in the ZnPc layer) for 20 s.

**Power-dependent PL measurements for Microcavities and control samples**

Power-dependent PL measurements at normal incidence were performed under 633 nm HeNe laser excitation using a 100× objective lens for microcavities, and a 40× objective lens for the control samples.

**Power-dependent TRPL measurements for Microcavities and control samples**

Transient photoluminescence (TRPL) was measured using a time-correlated single-photon counting (TCSPC) setup consisting of single-photon avalanche diodes (PDM series, Micro Photon Devices) and an Event Timer & TCSPC unit (PicoHarp 300). The $\lambda_c$ = 880 nm microcavity and control samples on fused silica substrates were excited by a supercontinuum laser (Fianium FemtoPower 1060, 40 MHz) with 630 nm excitation using a bandpass filter (FB630-10).

**Acknowledgements**

S.K.-C. was supported by the Natural Sciences and Engineering Research Council of Canada (NSERC) Discovery Grant program and the Canada Research Chairs Program. T.I. was supported by JSPS Overseas Research Fellowships and FRQNT Merit Scholarship Program (PBEEE). The authors benefit from their affiliation with the RQMP (https://doi.org/10.69777/309032) and wish to thank Prof. Joel Yuen-Zhou and Dr. Juan Pérez Sanchez for enlightening discussions at the early stages of this work.

**Figure**

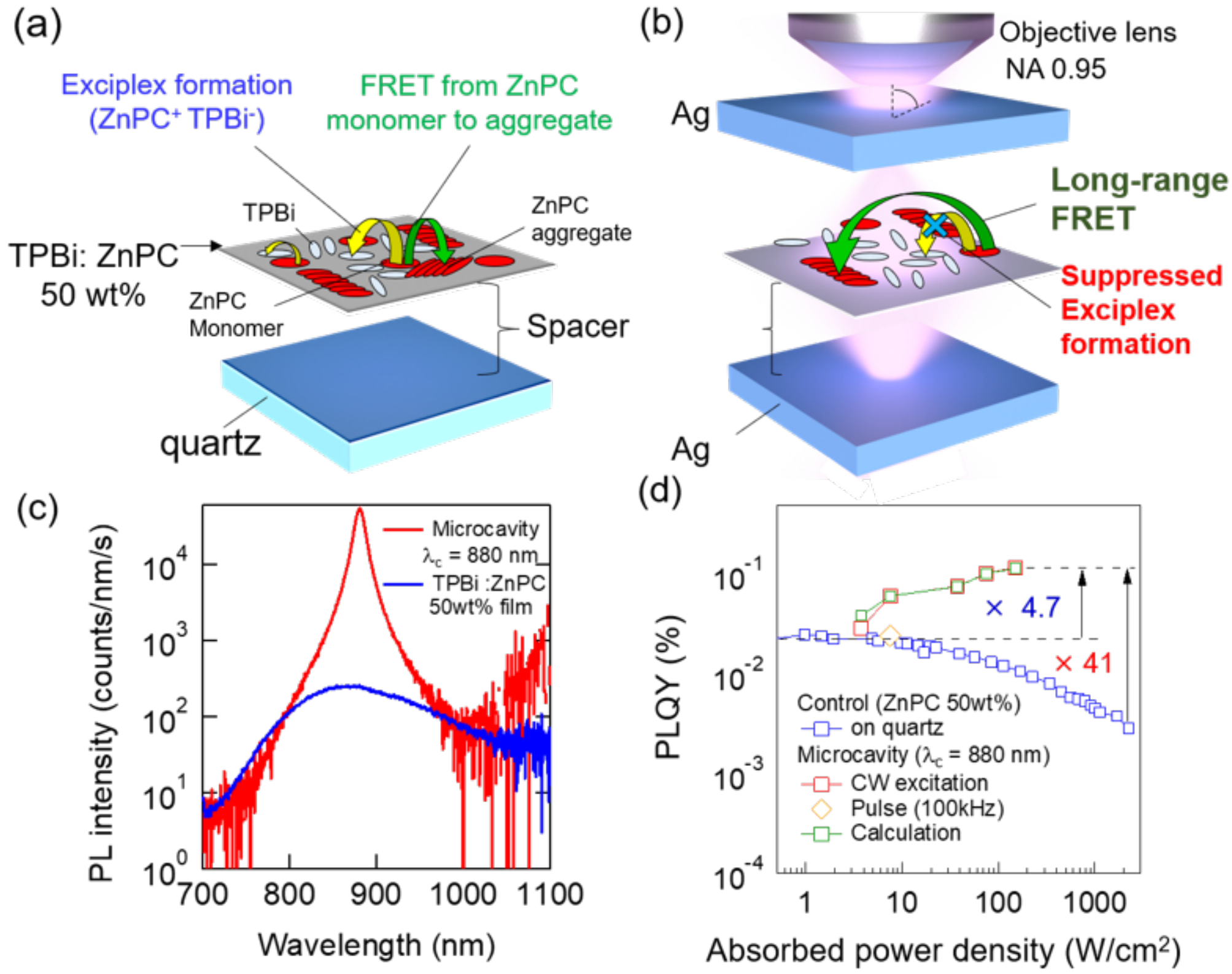

**Figure 1.** (a) Schematic showing exciplex formation via charge transfer from an excited ZnPc monomer (donor) to TPBi (acceptor). The FRET rate from ZnPc monomer to aggregates (~ 1.09 ± 0.09 ps) competes with charge transfer for exciplex formation (<200 fs). (b) Schematic showing long-range FRET from ZnPc monomer to aggregates inside a cavity. In the cavity, long-range FRET can be enhanced 400-fold compared with the control sample, suppressing exciplex formation and favoring excitation of aggregates. (c) PL spectrum of the microcavity and the control sample with TPBi: ZnPc 50wt% ($d = 5$ nm). (d) Absorbed power dependence of the PLQY for the $\lambda_C$ = 880 nm microcavity and the control sample on fused silica. The PLQY of the control was independently measured to be 0.028% using an integrating sphere (Table S1.1). The PLQY of the microcavity was measured under CW laser excitation at $\lambda$ = 633 nm and obtained by normalizing the number of photons collected within a numerical aperture (NA) of 0.95 to the number of absorbed photons (Fig.1b) and correcting for the out-coupling efficiency. (see Methods and Supplementary Information 2.7 for details). The yellow diamond plot represents the PLQY of the microcavity under excitation with a 100 kHz pulsed laser at λ = 633 nm. The green curve shows the PLQY simulated using a rate-equation model for the system (see Supplementary Information 2.4).

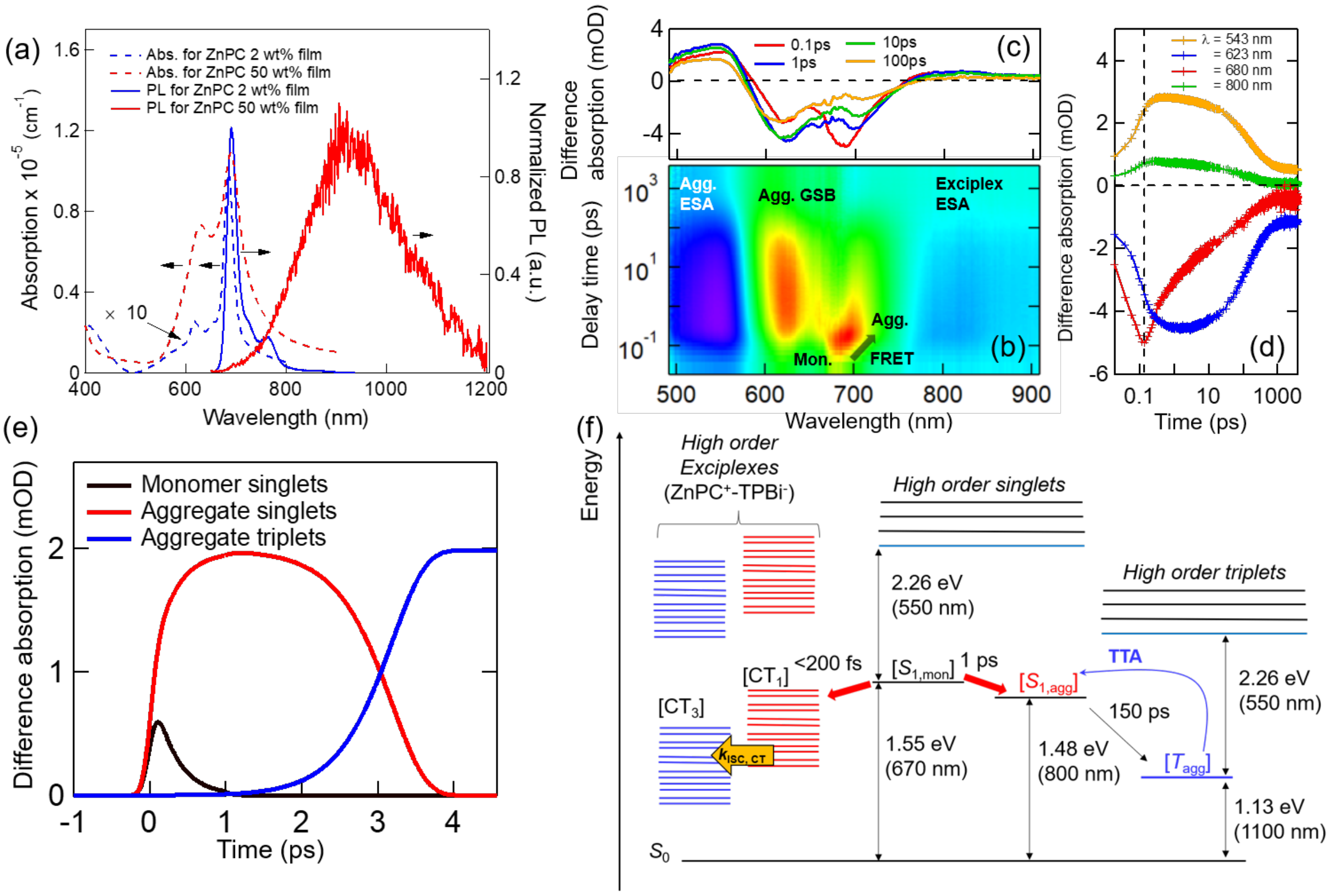

**Figure 2.** (a) Absorption and PL spectra of 200 nm-thick films of ZnPc doped in TPBi at 2 wt% and 50 wt% on fused silica substrates. (b) Transient absorption spectrum showing that after excitation, a bleach of ZnPc monomers at $\lambda = 680$ nm is observed, which evolves on a ~1 ps timescale into a bleach of ZnPc aggregates due to FRET. In addition, excited-state absorption (ESA) attributed to aggregate ZnPc appears near 550 nm, while ESA corresponding to exciplex ($ZnPc^{+}$-$TPBi^{-}$) is observed at wavelengths above 765 nm. (c) Transient absorption spectra of a TPBi:ZnPc 50 wt% doped film at 0.1 ps, 1 ps, 10 ps, and 100 ps following excitation at 680 nm. (d) Delay time dependence of difference absorption at $\lambda = 543$ nm, 623 nm, 680 nm, and 800 nm. (e) Global analysis results for kinetics of ZnPc monomer singlet, aggregate singlet, and aggregate triplets. The associated spectra are shown in Fig. S1.4. (f) Energy diagram of the TPBi:ZnPc 50 wt% film. $S_{1\text{mon}}$, $S_{1\text{agg}}$, $T_{\text{agg}}$, $CT_1$, and $CT_3$ correspond to the monomer singlet, aggregate singlet, aggregate triplet, exciplex singlet, and exciplex triplets, respectively. Energy diagrams for the TPBi:ZnPc 2 wt% and neat ZnPc films are shown in Fig. S1.10 and S1.13, respectively.

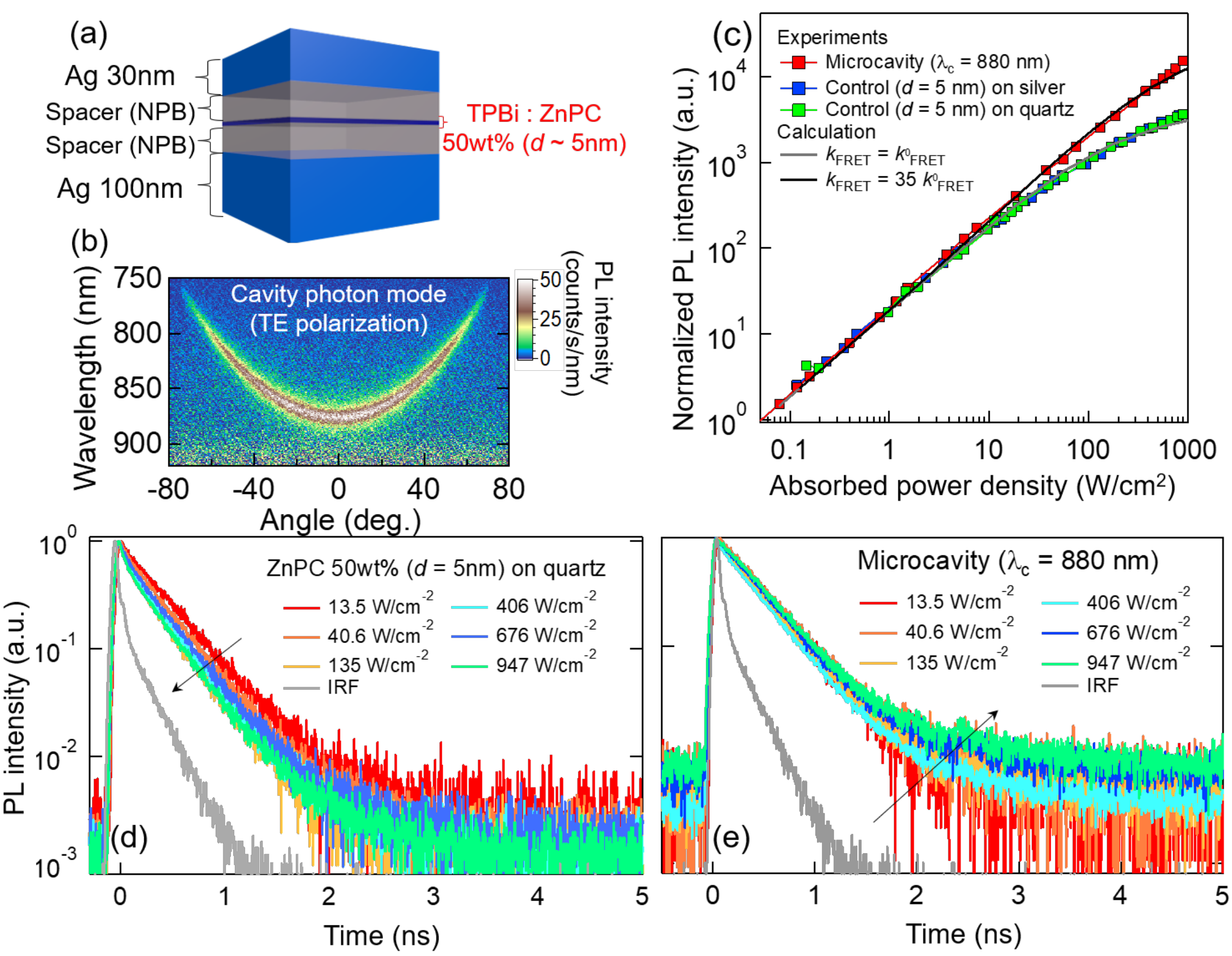

**Figure 3.** (a) Schematic representation of the microcavity structure consisting of Ag/TPBi/NPB/TPBi:ZnPc 50 wt%/NPB/Ag/Si substrate. Thickness for each layer showed in Table S2.1. (b) Angle-resolved PL for transverse electric (TE) mode for microcavities with $\lambda_c = 880$ nm. Similar figures for microcavities with $\lambda_C = 780$ nm, 860 nm, and 920 nm are provided in Fig. S2.12 and Fig. S2.13 for TE and TM polarization, respectively. (c) Absorbed power dependence of the PL intensity, measured at the normal incidence for the same microcavity. Similar figures for microcavities with $\lambda_C = 780$ nm, 860 nm, and 920 nm are provided in Fig. S2.11 and the corresponding PL spectra are shown in Fig. S1.15 and S2.10, respectively. The gray line shows the power dependence of the PL calculated by the kinetic model (see supplementary information 2.4). In contrast, the black line represents the calculation assuming that the FRET rate is enhanced by a factor of 35 $k_{FRET}{}^{0}$ denotes the FRET rate in the control sample. (d) (e) Time-resolved photoluminescence for the control sample on fused silica and the $\lambda_C = 880$ nm microcavity (see Methods for sample information). Note that the pump power was adjusted to match the absorbed power densities in the microcavity active layer to those of the control sample. Lifetime fits are shown in Figs. S1.14 and S2.5, respectively.

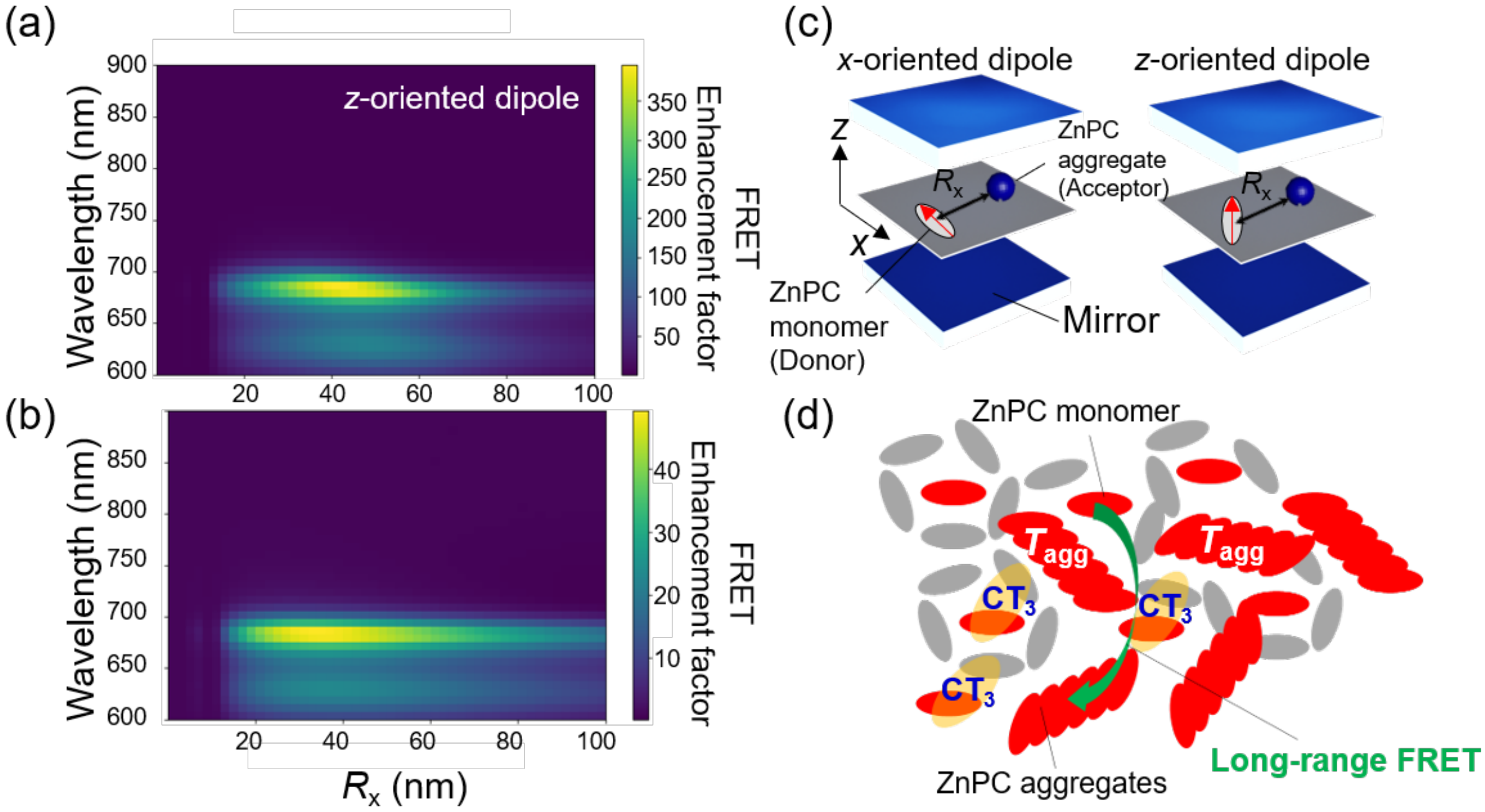

**Figure 4.** Contour plots of the FRET enhancement factor in the $\lambda_C$ = 880 nm microcavity as a function of $R_x$ and wavelength for (a) z-oriented dipole, and (b) x-oriented dipole. The long-range FRET enhancement factors reach approximately 400 and 50 for the *z*- and *x*-oriented dipoles, respectively. (c) Schematic representation of a system consisting of a donor (ZnPc monomer) and an acceptor (ZnPc aggregate) with displacements $R_z$ and $R_x$ along the z- and x-axes, respectively. (d) Schematic illustration showing that, in a high exciton density regime, aggregate and exciplex triplets can occupy nearby aggregate- and CT- forming sites, thereby reducing the effective rates of both short-range FRET and charge transfer, while accelerating long-range FRET from ZnPc monomers to aggregates.

· **REFERENCES**